# COMPOSITE DESIGN PATTERN FOR FEATURE-ORIENTED SERVICE INJECTION AND COMPOSITION OF WEB SERVICES FOR DISTRIBUTED COMPUTING SYSTEMS WITH SERVICE ORIENTED ARCHITECTURE

Vishnuvardhan Mannava[1] and T. Ramesh[2]

[1]Department of Computer Science and Engineering, K L University, Vaddeswaram, 522502, A.P., India
`vishnu@kluniversity.in`
[2]Department of Computer Science and Engineering, National Institute of Technology, Warangal, 506004, A.P., India
`rmesht@nitw.ac.in`

### ABSTRACT

*With the advent of newly introduced programming models like Feature-Oriented Programming (FOP), we feel that it will be more flexible to include the new service invocation function into the service providing server as a Feature Module for the self-adaptive distributed systems. A composite design patterns shows a synergy that makes the composition more than just the sum of its parts which leads to ready-made software architectures. In this paper we describe the amalgamation of Visitor and Case-Based Reasoning Design Patterns to the development of the Service Invocation and Web Services Composition through SOA with the help of JWS technologies and FOP. As far as we know, there are no studies on composition of design patterns for self adaptive distributed computing domain. We have provided with the sample code developed for the application and simple UML class diagram is used to describe the architecture.*

## 1. INTRODUCTION

Patterns are reusable good-quality design practices that have proven useful in the design of software applications. A large number of software design patterns have been described, catalogued and included in software tools. Design patterns yields better-quality software within reduced time frames. When designing software two or more patterns are to be composed to solve a bigger problem. Pattern composition has been shown as a challenge to applying design patterns in real software systems [4]. Composite patterns represent micro architectures that when glued together could create an entire software architecture. Thus pattern composition can lead to ready-made architectures from which only instantiation would be required to build robust implementations. A composite design patterns shows a synergy that makes the composition more than just the sum of its parts. As far as we know, there are no studies on composition of design patterns for self adaptive distributed computing domain.

Composition of web services is the most important technique that the developers are now a day's most interested in providing the web services with. So in order to achieve the composition in our paper we are adopting to the Service Oriented Architecture (SOA). SOA provides composition of the services provided by the different servers to fulfil the complex service requests of the clients. The requester's of the web services will request a functionality that cannot be fulfilled by a single service provider. In many cases the web services are having limited functionality. So when a client requests for a complex service, he many not find a WSDL that provides the perfect execution of the service. In that case in order to fulfil the users request that require multiple tasks may fail due to unavailability of suitable Web services WSDL xml file. In such scenario there is a requirement to compose the web services to satisfy the requester's complex request.

We can use either Java or .Net platform to develop the SOA with web services. Even if you use any one of them it's the web services that make the SOA compatible with the composition of the services in any platform. When using the web services, interface definitions are provided using the WSDL. Each service at the server is associated with a WSDL document. WSDL is expressed using XML. It defines the input and output parameters of a Web service in terms of XML Schema. So with the help of the Simple Object Access Protocol (SOAP) is used as XML-based protocol for exchanging information in a distributed environment. SOAP provides a common message format for exchanging data between clients and services. The SOAP messages can be exchanged with the help of HTTP GET, POST, PUT requests. The HTTP GET is used to request the WSDL messages. The HTTP POST used for the service request/ response.

A web service is defined as an interface which implements the business logic through a set of operations that can be accessed with the help of standard Internet protocols [7]. In this paper we will propose a design pattern that is an amalgamation of the Visitor pattern [6] and Case-Based Reasoning pattern [11]. A design pattern is a particular form of recording information about a design such that the same pattern can be applied in future, if same situation is repeated to solve problem . So the design patterns are accepted in wide range of object-oriented designs. Collections of design patterns can be found in numerous publications.

We will use the case-based reasoning pattern for the purpose of decision making at the client side. When the client asks for a service, this pattern is helpful in deciding which among the available Web Service Description Language (WSDL) will be helpful to perform the required task. Initially the client will request all the web service providers to send the WSDLs of their respective services. Then the pattern in the client will decide the best WSDL of a service provider to be selected according to the case based selection of the service with respect to the WSDL that can be used. The visitor pattern at the server side will provide the services to the clients request by either just invocation of the service or by the composition of the services at different servers.
In this proposal we have also emphasized the Feature-Oriented Programming (FOP) based Service injection into the current service provider[8]. The FOP [2] has been the most popular programming model that has showed a vast degree of enhancements in the field of Software Product-Lines (SPL). So using the feature based functional module insertion we can include the new services by maintaining transparency to the user of the services in run time.

## 2. RELATED WORK

In this section we present some works that deal with different Design patterns oriented work. There are number of publications reporting the reusability features of the design patterns when the same problem occurs in the future.

Demian Antony D'Mellon, V.S. Ananthanarayana, and Supriya Salian in paper [3] proposed the review of Web Services composition Architectures and techniques used to generate new services.

V.S.Prasad Vasireddy, Vishnuvardhan Mannava, and T. Ramesh paper [10] discuss applying an Autonomic Design Pattern which is an amalgamation of chain of responsibility and visitor patterns that can be used to analyze or design self-adaptive systems. They harvested this pattern and applied it on unstructured peer to peer networks and Web services environments.

Olivier Aubert, Antoine Beugnard [9] they proposed an Adaptive Strategy Design Pattern that can be used to analyze or design self-adaptive systems. It makes the significant components usually involved in a self-adaptive system explicit, and studies their interactions. They show how the components participate in the adaptation process, and characterize some of their properties.
Don Batory, Jacob Neal Sarvela, and Axel Rauschmayer [2] in this work they have explained that Step-wise refinement is a powerful paradigm for developing a complex program from a simple program by adding features incrementally. They presented the AHEAD (Algebraic Hierarchical Equations for Application Design) model that shows how step-wise refinement scales to synthesize multiple programs and multiple noncode representations.

In Vishnuvardhan Mannava, and T. Ramesh paper [13] they have proposed a design pattern for Autonomic Computing System which is designed with Aspect-oriented design patterns and they have also focused on the amalgamation of the Feature-oriented and Aspect-oriented software development methodology and its usage in developing a self-reconfigurable adaptive system.
In Vishnuvardhan Mannava, and T. Ramesh paper [14] they have proposed a system for dynamically configuring communication services. Server will invoke and manage services based on time stamp of service. The system will reduce work load of sever all services in executed by different threads based on time services are executed, suspended and resumed.

In Vishnuvardhan Mannava, and T. Ramesh paper [15] they have proposed an adaptive reconfiguration compliance pattern for autonomic computing systems that can propose the reconfiguration rules and can learn new rules at runtime.

Because of the previous proposed works as described above, we got the inspiration to apply the aspect oriented design patterns along with inclusion of the feature-oriented software development capability to autonomic systems.

## 3. PROPOSED AUTONOMIC DESIGN PATTERN

In this paper we have proposed a design pattern that provides the composition of the web services by the support of service Oriented Architecture (SOA) [7]. Here we are using two design patterns both of them results to an amalgamation of our proposed design pattern. Initially when the user requests for the service the client machine will generate the HTTP GET request to all the service providing servers in the network to send the respective WSDL (Web Service Description Language). Then after that the Case-Based Reasoning Design Pattern [11] at the Client side machine will use the WSDLs that it have received from different servers in the network and also the service request that was given as input to the client by the user to decide the Best perfect matched web Service provider to the user's request. All the above steps of decision making will be performed on the bases of some cases or statement based decision making, so that the WSDL that matched the service request that is requested by the user can be used to fulfil the service.
Then once a perfect WSDL has been selected based on the condition that the input parameters of the service requested have to match the input parameters in the WSDL XML file, then the SOAP message is used to send the request to the Server.

On the other hand that is at the server side the request for the service is received in the form of SOAP Message by the server. Then it will use the Visitor Design Pattern [6] to handle the request. If the requested service is available at the server, then it will provide service without any problem. If the service is a complex service then the server will take help of other servers that provide the service which is not available by itself. So in order to provide the Composition of the services at different servers the current server that knows where the service is available will send the service request to the peer server. The current server have to know the access reference to the operation at the other server to invoke that service and perform the clients requested complex task. For this purpose we have used the Visitor pattern in order to get the reference to access the other server's services and to invoke the service operation in the peer server.

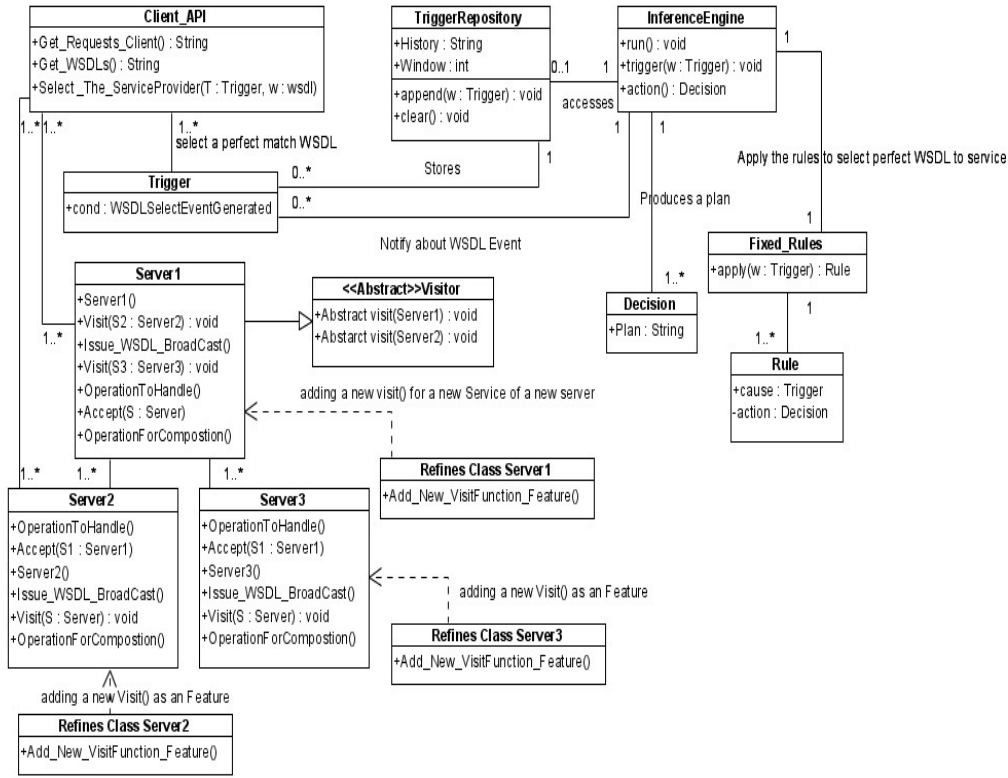

Figure 1: Applying Design Pattern for the Distributed Computing System

In this way the server that is only having the worth to serve the two functionalities out of three in the requested task at that situation will take help of the other peer servers to fulfill the job it has accepted from the client. Here the server can handle multiple clients in the same time using the thread based processing of the requests. Server can allocate a separate thread for each and every client to handle the requests concurrently.

We have provided a capability to our proposed design pattern that, it can add a new service operation to a server by just including the new service as a Feature Module with the help of Feature-Oriented Programming (FOP). Here the injection of the new service is done without disturbing the already existing code or writing again from starch is to develop the application is eliminated[8].

So in our design pattern we have provided the capabilities of Web Services Composition, Service Invocation, and Inclusion of the new Service Operation as a Feature Module, all the three are provided in each and every server in a Distributed Application.

## 4. DESIGN PATTERN TEMPLATE:

To facilitate the organization, understanding, and application of the proposed design patterns, this paper uses a template similar in style to that used in [11]. Likewise, the Implementation and Sample Code fields are too application-specific for the Design Patterns presented in this paper.

### 3.1. Pattern Name:

Feature-Oriented Service Injection and Composition of Web Services with SOA

### 3.2. Classification:

Structural –Decision Making

### 3.3. Intent:

Systematically applies the Design Patterns to a distributed Computing System to provide the Web Services Composition and service injection with a Refinement class for providing new service in the servers in terms of a Feature Module.

### 3.4. Context:

Our design pattern may be used when:

- The service requested is a complex service and need to be executed with the help of collaboration of different servers in the network
- To include the new service operations into servers as a Feature Modules [2].
- When the service in the other server need to be accessed, and for that purpose we can use the use this pattern to provide the Invocation capability to our distributed Application.
- In order to provide the Composition of the services, we may use this pattern which provides the SOA (Service Oriented Architecture) based composition with the JWS (Java Web Services)[7].

### 3.5. Proposed Design Pattern Structure:

A UML class diagram for the proposed design Pattern can be found in Figure 1.

### 3.6. Participants:

a) **ClientAPI:** It is Responsible for the purpose of accepting the requests from the Client and then requesting the servers in the network that are responsible for providing the services will send the WSDLs of the respective Web Service providers. Once the above two tasks are completed then it will generate a Trigger Event to find the perfect WSDL that can serve the user requested service.

b) **Trigger:** It is responsible for the purpose of accepting the events that are generated by the clientAPI to check whether the selected WSDL of a web service provider is suitable to handle the client's request.

c) **Inference Engine:** It is responsible for accepting the events from the Trigger and then checking the WSDL with respect to the service that is requested by the client based on the some case based Reasoning statements or rules provided in the FixedRules class.

d) **FixedRules:** This will take the WSDL and then it will apply some rules to get some inference results from the operation to check whether the WSDL can handle the service requested by the client.

e) **Decision:** Here based upon the results from the FixedRules class it will help the InferenceEngine to select a perfect WSDL by applying or checking the Service requested by client and results to take decision whether it can handle the request or not.

f) **TriggerRepository:** it will look after the backup storage or just for storing the cause, situation, and results of the particular Event for the future based analysis and enquire.

g) **ServerN(N=1 or 2 or 3):** It is the web service provider which will look after providing the services to the clients who have requested them. Here we are using the visitor design pattern, which is very important aspect in our whole paper. Because it is responsible for invoking the services at the servers and also if a service is not available for current instance of a situation and the server need that service to complete the complex task of a client, so for that purpose it can invoke the service that is provided by the neighbouring server with the help of visitor pattern. Visitor pattern will get the reference of the peer server to access or invoke the service in the peer server and then retrieve the result from it to fulfil the client's request. The web service composition is provided with the help of Service Oriented Architecture (SOA).

h) **Refines Class ServerN(N=1 or 2 or 3):** It is responsible for the purpose injection of the new service providing servers Visit() as a Feature Module using the Feature-Oriented Programming (FOP)[1].

The view of our proposed design pattern for the Distributed Computing System can be seen in the form of a class diagram see Figure 1.

The flow of control in the Distributed Computing System can be shown with a sequence diagram in Figure 2.

### 3.7. Consequences:

a) With the help of the Web Service Description Language (WSDL) all the clients can get information about the services that are currently provided by the peer servers.
b) We use the Feature-Oriented Programming [2][1] to insert the new service providers visit (Reference Object) into current running server as a feature into the current executing services code.
c) With the help of Service Oriented Architecture (SOA) for the purpose of web service composition using the JWS (java web services) we can provide the composition of the service in a distributed application environment.
d) Invocation of the service can be done with the help of the visitor design pattern.

### 3.8. Related Design Patterns:

a) **Worker Object Pattern [12]:** The worker object pattern is an instance of a class that encapsulates a worker method. A worker object can be passed around, stored, and invoked. The worker object pattern offers a new opportunity to deal with otherwise complex problems. It will provide the server with the facility to handle the service request form different clients in a separate per client connection. We may use this pattern in different situations like, implementing thread safety in swing applications and improving the responsiveness of the UI applications to performing authorization and transaction management.

b) **Strategy Design Pattern [4]:** This pattern can be used to define a family of algorithms, encapsulate each one, and make them interchangeable. Strategy lets the algorithm vary independently from the clients that use it.

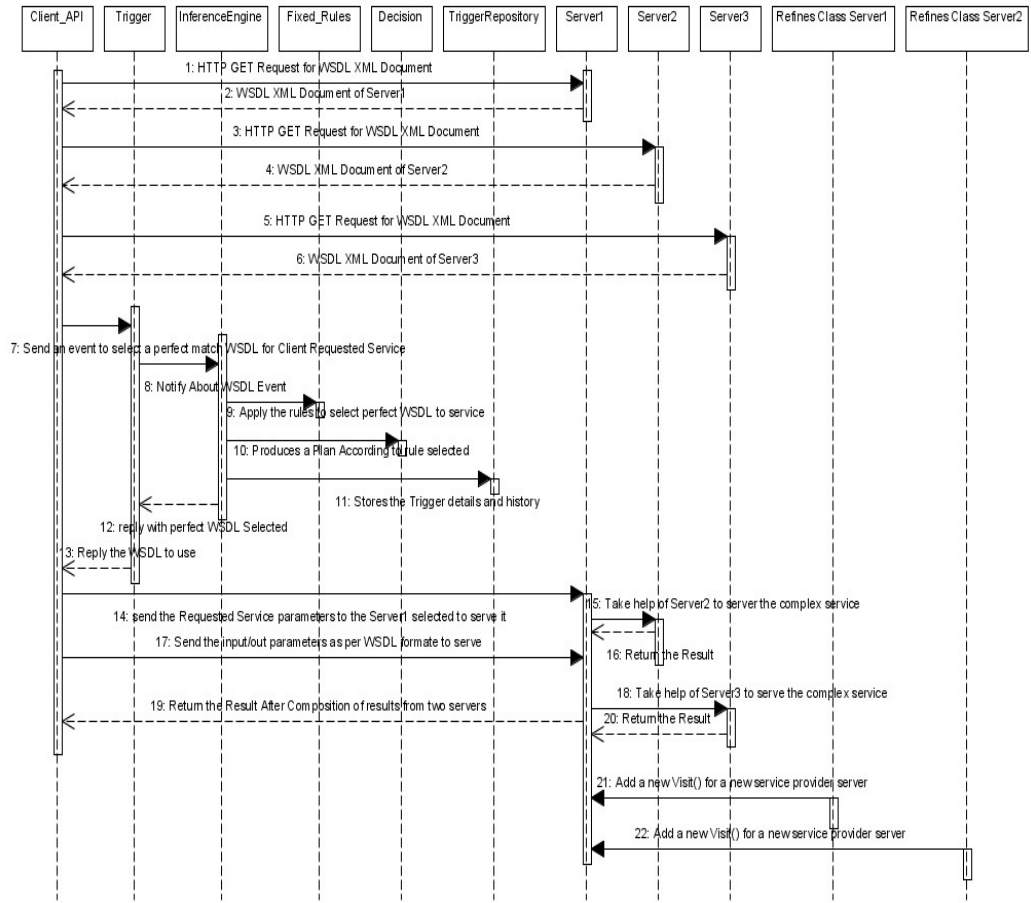

Figure 2: Sequence Diagram for the proposed Design Pattern

## 3.9 Role of our design pattern in Autonomic computing system

a) **Case-Based Reasoning Design Pattern [11]:** The Case-Based Reasoning Design Pattern in will apply the rule based decision making mechanism to determine a correct reconfiguration plan. This design pattern will separate the decision-making logic from the functional logic of the Application. In our proposed pattern we will use this pattern for the decision making purpose to decide which WSDL among the received WSDLs from the service providers can fulfill the clients request in accordance to the service requested by the client.

b) **Visitor Design Pattern [5]:** The visitor pattern turns the tables on OO model and creates an external class to act on data in other classes. We use this pattern when we want to perform an operation on the data contained in a number of objects that have different interfaces.

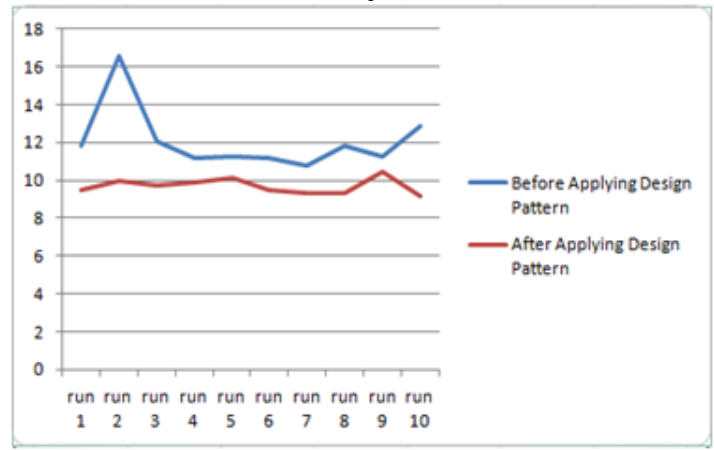

Figure 3: Profiling data after applying proposed AOP pattern and before applying with OOP pattern

## 5. PROFILING RESULTS

We are presenting the profiling results taken for ten runs without applying this pattern and after applying this pattern using the profiling facility available in the Netbeans IDE. The graph is plotted taking the time of execution in milliseconds on Y-axis and the run count on the X-axis. The graph has shown good results while executing the code with patterns and is shown in Figure 3.This can confirm the efficiency of the proposed pattern.

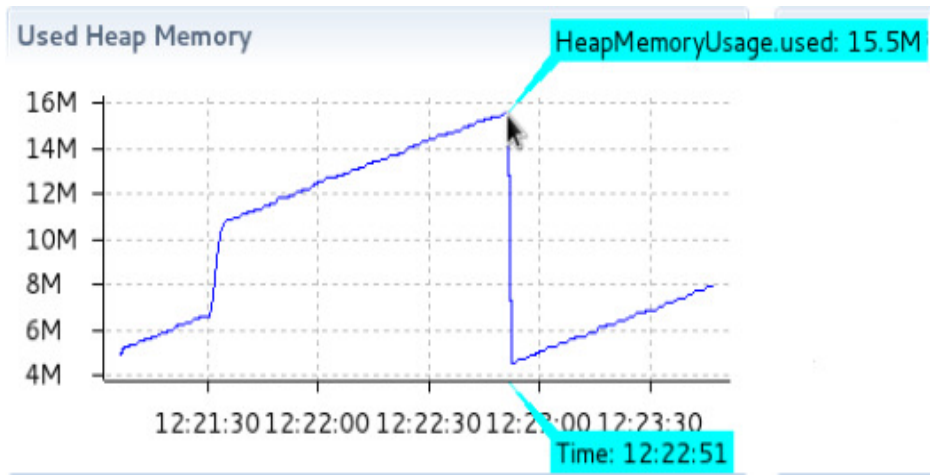

Figure 4: The Heap Memory usage after applying Aspect-Oriented Programming Techniques to Visitor pattern

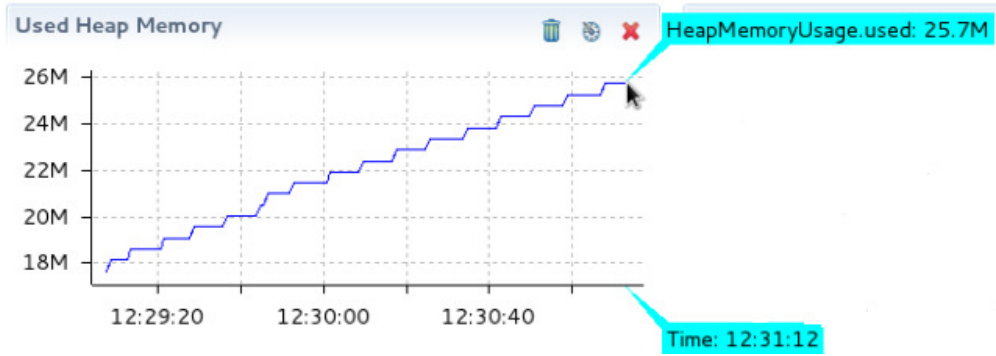

Figure 5: The Heap Memory usage after applying Object Oriented Visitor Design Pattern

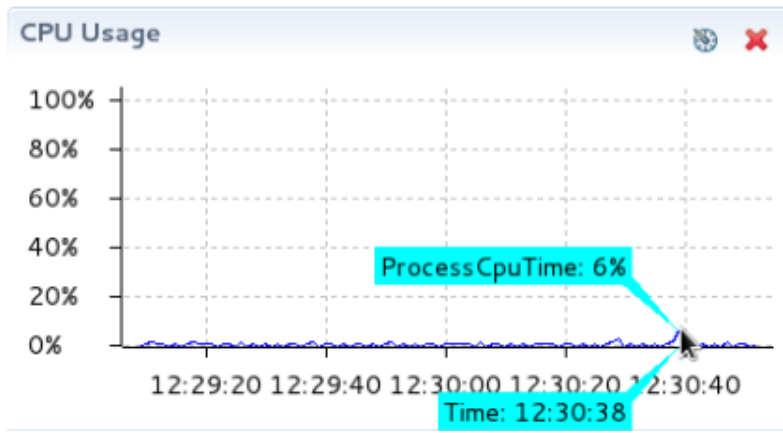

Figure 6: The CPU usage time after applying Aspect-Oriented Programming Techniques to Visitor pattern

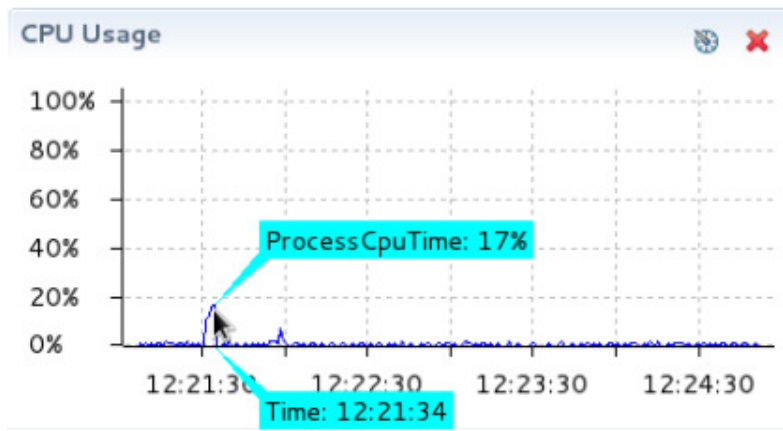

Figure 7: The CPU usage time after applying Object Oriented Visitor Design Pattern

### 5.1. Simulation Results after Visitor Pattern

In our implementation we can evaluate the effectiveness of our implemented case-study with SOA and Socket Programming. In order to make our proposal clear we have successfully

International Journal of Web & Semantic Technology (IJWesT) Vol.3, No.3, July 2012developed some critical parts of our system i.e., implementation of visitor pattern integrated with SOA with the help of Aspect-Oriented Programming (AOP) and new service injection as a Feature with the help of Feature-Oriented Programming (FOP), and at the same time we have implemented the same pattern with Object-Oriented visitor design pattern with pure socket programming.

The simulation results for the code developed to prove the benefits of Aspect Oriented Programming features are collected with respect to:

- Used Heap memory
- Process CPU Time

### 5.2. Discussion

From the Figure 4 and Figure 5 we can evaluate that the amount of Heap Memory used by applying Object Oriented Programming Visitor pattern is 51492 kbytes and where as for the amount of Heap Memory used with AOP Visitor pattern along with SOA is 10,691 kbytes. It's clear that the application developed using Aspect-Oriented Programming along with SOA takes less heap memory when compared to implementation with respect to Object Oriented Visitor design pattern.

From the Figures 6 and Figure 7 we can evaluate that the amount of CPU Time used by applying AOP Visitor pattern along with SOA is 7.350sec and where as for the amount of CPU Time used by applying Object Oriented Programming Visitor pattern is 10.270 sec. It's clear that the application developed using Aspect-Oriented Programming takes less CPU Time when compared to implementation with respect to OOP based Visitor design pattern.

## 6. CONCLUSION

In this paper we have proposed a pattern to facilitate the ease of Design Pattern for Feature-Oriented Service Injection and Composition of Web Services with SOA. So with the help of our proposed design pattern named Service Injection Design pattern for Distributed computing systems, provide services to clients with the help of Object-Oriented design patterns. So with this pattern we can handle the service-request of the clients and inject the new services into server's code as feature modules. We have also focused on how we have achieved the Composition of the Web services through Service Oriented Architecture (SOA). The Several future directions of work are possible. We are examining how the SOA Oriented Design patterns can be used to provide much powerful application developments in distributed environments.

**A. INTERFACES DEFINITION FOR THE COMPOSITE DESIGN PATTERN ENTITIES**

Some of the Interfaces for the classes are provided as below:
**Server1 Sample Code (this is same for both server 2 and 3)**
```
Public Abstract class Visitor
{
   Public abstract void visit (Server s2);
   Public abstract void visit (Server s3);
}
Public class Server1 extends Visitor
{
Boolean value;
int date
int i;
```

```
Float price;
Public Server1 ()
{
   Value=false;
}
Public void visit (Server s2)
{
//code logic for providing service of Engineering books.
// get the dates that are expected to be delivered on.
//set the value [i++]=True for specifying that the books requested are in stock.
}
Public void visit (Server s3)
{
//code logic for providing service of Medical books.
//get the dates that are expected to be delivered on.
//set the value [i++]=True for specifying that the books requested are in stock.
}
.. //add as many service providers according to your requirement.
Public int getThe Delivery&priceDetails()
{
//Code logic for providing the details or reply the status of the requested books and there price.
}
.. //more business logic if required can be added.
}
```

**The main class in the server**
```
Server1 s1 =new Server1();
String Result;
Run()
{
servicesToHandle=getcount();
Server2.accept (s1);
Server3.accept(s2);
}
Result=S1.getTheDelivery\&priceDetails();

Server2 or Server3  //business logic
Public class Server2 or 3
{
//initialization through constructor
//business logic that is required to handle the requested service execution.
Public sendWSDLDoc()
{
}
Public string EngBooksSearch()
{
        //required coding logic
}
Public String MedicalBooksSearch()
{
        //required coding logic
}
.. //More number of service providers functions.
Public void accept(Visitor V1)
{
V1.visit(this);
}
}
```